\documentclass[twocolumn,prl,amsmath,amssymb,superscriptaddress]{revtex4}

\usepackage{graphicx}
\usepackage{dcolumn}
\usepackage{bm}

\def\reff#1{(\ref{#1})}

\begin{document}

\title{Artificial magnetism and left handed media from dielectric rings and rods}

\author{L. Jelinek}
\email{l_jelinek@us.es} 
\affiliation{Departamento de Electr\'onica y Electromagnetismo, Universidad de Sevilla, 41012-Sevilla, Spain}

\author{R. Marqu\'es}
\email{marques@us.es} 
\affiliation{Departamento de Electr\'onica y Electromagnetismo, Universidad de Sevilla, 41012-Sevilla, Spain}

\date{\today}

\begin{abstract}
It is shown that artificial magnetism with relatively large frequency bandwidth can be obtained from periodic arrangements of dielectric rings. Combined with dielectric rods, dielectric rings can provide 3D isotropic left-handed metamaterials being an advantageous alternative to metallic split ring resonators and/or metallic wires when un-detectability by low frequency external magnetic fields is desired. Furthermore it is shown that dielectric rings can be also combined with natural plasma-like media to obtain a left-handed metamaterial.
\end{abstract}


\maketitle


Metamaterials, that is, artificial effective media with properties not found in Nature, such as negative magnetic permeability and/or permitivity, have been recently the subject of a big wave of scientific interest due to its unique new physical properties and promising applications (see, e.g. \cite{Marques} and references therein). At the present state of the art, the most common way to artificial magnetism uses metallic rings loaded either by a lumped capacitor \cite{Shelkunoff} of by a distributed capacitance \cite{Pendry-1999}. This last configuration - the so called Split Ring Resonator (SRR) - is by far the most common configuration due to its easy manufacturing by means of standard photo-etching techniques. There can be, however, many other possibilities for obtaining artificial magnetism, and, in particular, an intensive research has been recently aimed to the substitution of metallic SRRs by purely dielectric resonators in order to reduce losses and/or minimize interactions with low frequency external magnetic fields. Among the most promising proposals in this direction are high permittivity dielectric spheres working at Mie resonance \cite{Holloway-2003,Vendik-2006} and square-shaped dielectric resonators \cite{Popa-2008}. Along this letter we will show that uniform dielectric rings (DRs) can also can provide strong artificial magnetism. 

To show the principle of operation of the DR, let's assume a ring of mean radius $a$ made of a ``dielectric wire'' of radius $b$, as it is sketched in Fig. 1a. Let the permittivity of the ring be $\varepsilon  = \varepsilon '\left( {1 - j \, \text{tg}\delta } \right)$ and let the ring be placed in a homogeneous time varying magnetic field directed along the ring axis. For high enough permittivities, $\varepsilon$, it can be assumed that the electric field is strongly confined inside the ring due to the boundary condition ${\bf E}\cdot{\bf n} = \varepsilon_0/\varepsilon{\bf E_0}\cdot{\bf n} \approx 0$ (see Fig. 1a). Therefore, assuming an uniform distribution of the electric field inside the ring, an axial magnetic field will induce an uniform displacement current in the ring given by
\begin{equation}
  \label{eq1}
  I = j\omega \varepsilon AE_\varphi,
\end{equation}
where $A=\pi \, b^{2}$ is the cross-section of the wire and $E_{\varphi}$ is the angular component of the electric field, which can be determined from Faraday's law 
\begin{figure}
\centering
\includegraphics[width=0.9\columnwidth]{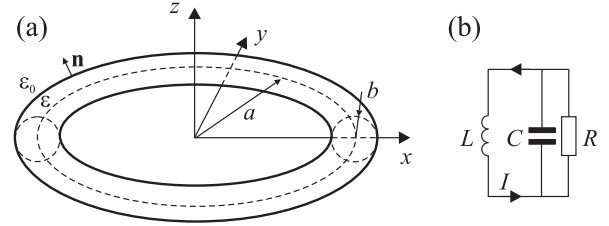}
\caption{\label{Fig1} (a) Sketch of the proposed dielectric particle (b) Equivalent circuit model.}
\end{figure}
\begin{equation}
  \label{eq2}
  2\pi aE_\varphi   =  - j\omega \left( {\phi ^{\operatorname{\text{int}} }  + \phi ^{\text{ext}} } \right),
\end{equation}
where $\phi ^{\operatorname{\text{int}} }  = LI$ is the internal magnetic flux due to the self inductance of the ring and $\phi ^{\text{ext}} $ is the external driving magnetic flux. Substituting (1) into (2) we obtain
\begin{equation}
  \label{eq3}
  \left( {\frac{{2\pi a}}{{j\omega \varepsilon A}} + j\omega L} \right)I =  - j\omega \phi ^{\text{ext}}.
\end{equation}
In (3) the first term in the bracket can be easily recognized as the impedance $Z=(1/R+j\omega C)^{-1}$ coming from the parallel combination of a capacitor $C$ and a resistor $R$
\begin{equation}
  \label{eq5}
 C = \frac{{\varepsilon 'A}}{{2\pi a}}\;\;;\;\;\; R = \frac{{2\pi a}}{{A\omega \varepsilon '\text{tg}\delta }}
\end{equation}
which actually are the capacitance and the resistance of an ideal parallel plate capacitor of plate surface equal to the wire cross-section and distance between plates equal to the circumference of the ring. Equation \reff{eq3} leads to the equivalent circuit shown in Fig. 1b, where $L$ can be evaluated as \cite{Landau-8} 
\begin{equation}
  \label{eq6}
  L = \mu _0 a\left[ {\ln \left( {8a/b} \right) - 7/4} \right].
\end{equation}
Equation \reff{eq3} can also be used to calculate the axial magnetic polarisability of the particle 
\begin{equation}
  \label{eq7}
  \alpha^{mm}_{zz}  = \frac{{\left( {1 - j \, \text{tg}\delta } \right)\left( {\pi a^2 } \right)^2 /L}}
{{\left( {\frac{{\omega _0^2 }}
{{\omega ^2 }} - 1 + j \, \text{tg}\delta } \right)}},
\end{equation}
where $\omega _0  = 1/\sqrt {LC} $. Therefore, the electrical size of the DR at resonance is
\begin{equation}
  \label{eq8}
\frac{{2a}}
{{\lambda _0 }} = \frac{{\sqrt 2 }}
{\pi } \cdot \frac{{a/b}}
{{\sqrt {\left[ {\ln \left( {8a/b} \right) - 7/4} \right]} }} \cdot \frac{1}
{{\sqrt {\varepsilon '_r } }}.
\end{equation}
From \reff{eq8} it follows that in order to reduce the electrical size of the DR, high permittivity dielectrics and/or low $a/b$ ratios must be used. Since losses should increase with permittivity, it will be advantageous to use $a/b$ ratios as small as possible. Assuming $a/b = 3$, we can see that in order to achieve an electrical size  $a/\lambda_0\,\sim\,0.1$, the real part of the ring permittivity must be $\varepsilon\,\sim\, 100\,\varepsilon_0$, which is a quite achievable value at microwave and THz frequencies.
\begin{figure}
\centering
\includegraphics[width=0.9\columnwidth]{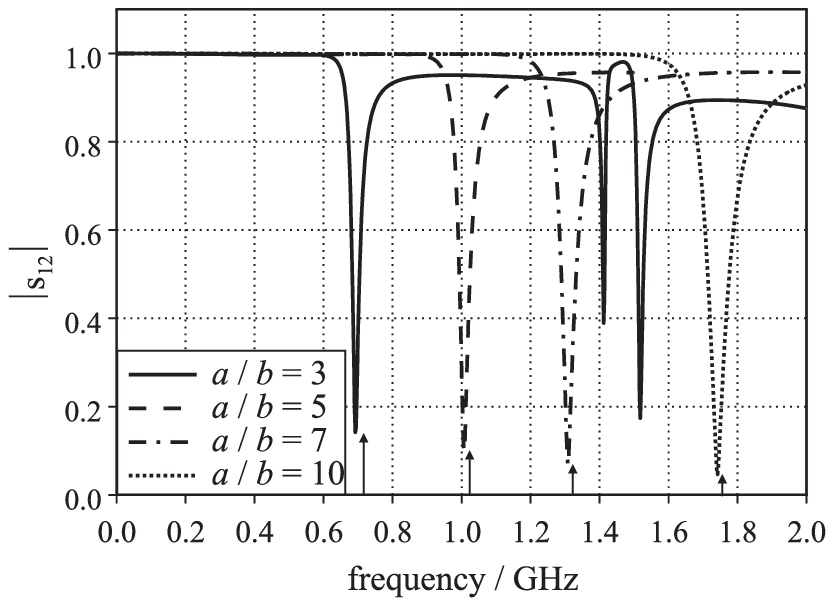}
\caption{\label{Fig3} Transmittances through a square TEM waveguide of lateral size $4a$ loaded by a dielectric ring resonator with $a=\text{15 mm}$ and $\epsilon_{r}=250$ for several $a/b$ ratios. Curves come from numerical simulations. Arrows show the resonances predicted by the reported analytical model.}
\end{figure}
\begin{figure}
\centering
\includegraphics[width=0.9\columnwidth]{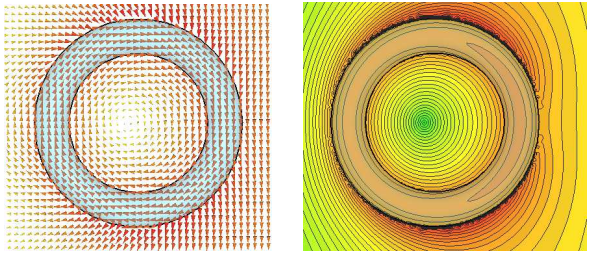}
\caption{\label{Fig2} (Color Online) Vector and contour plot of the electric field intensity in the vicinity of the dielectric ring resonance. The incident wave is the TEM waveguide mode with the magnetic field along the axis of the ring. The contour plot shows the absolute value of the field. The ring parameters are the same as in Fig. 2.}
\end{figure}

To check our analytical results, several DRs with different $a/b$ ratios have been designed and simulated using CST Microwave Studio. To obtain the resonance frequency and the field distribution at resonance an ideal square TEM waveguide (with upper and bottom perfect electric conducting walls, and lateral perfect magnetic conducting walls) was loaded with a dielectric ring, and the scattering parameters were calculated. The dielectric ring was placed at the center of the TEM waveguide, with its axis along the incident magnetic field. With this configuration, any resonance of the dielectric ring will appear as a sharp dip in the transmission coefficient. Figure 2 shows the simulated transmission coefficients and the resonances predicted by our analytical model for several values of the $a/b$ ratio. A very good agreement can be observed, even for $a/b$ ratios as smaller as $a/b = 3$. The higher order DR resonances which appear in the figure for the smaller $a/b$ ratio are electric resonances, whose description is outside the scope of our model. The electric field simulations shown in Fig. 3 also confirms the hypothesis underlying the model, particularly the uniformity of the field along the ring.
Besides the dielectric nature, a noticeable advantage of the proposed DR is that it exhibits all planar symmetries needed for isotropic metamaterial design \cite{Baena-2007}. Such isotropic composite, whose unit cell is shown in the inset of Fig. 4, can be described by the homogenization model developed in \cite[Eq. 13]{Baena-2008} combined with the magnetic polarizability (\ref{eq7}) which leads to the effective magnetic permeability shown in Fig. 4. The relative bandwidth of the negative permeability region is about $10\,\%$, one order of magnitude larger than other previously reported bandwidths for dielectric negative permeability media \cite{Vendik-2006,Popa-2008}. The reason for this enhanced bandwidth may be the strong confinement of current in the perifery of the DR, which results in a stronger magnetic moment.

\begin{figure}
\centering
\includegraphics[width=0.9\columnwidth]{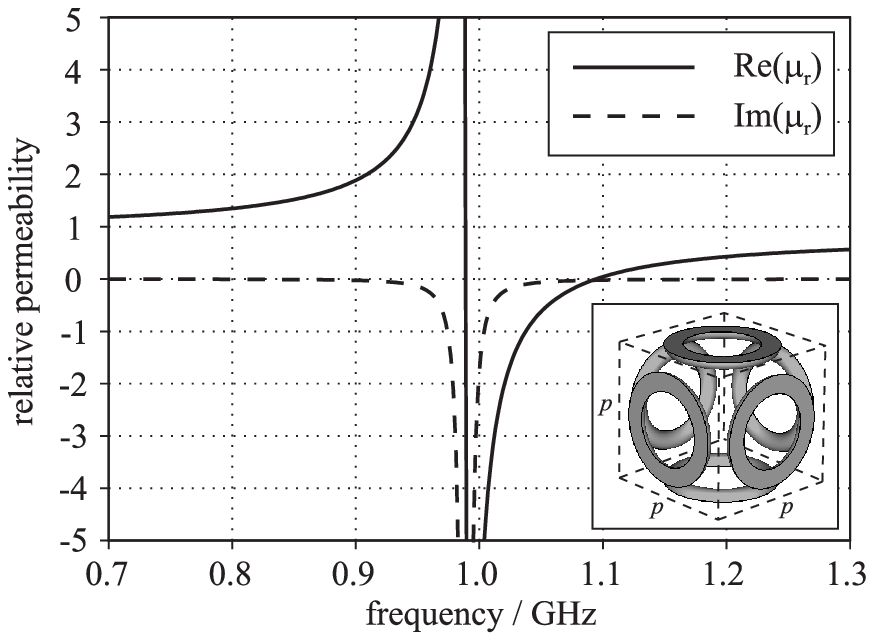}
\caption{\label{Fig4} Effective permeability of an isotropic magnetic medium composed of DRs. Geometrical ring parameters are: $a=\text{15 mm}$ and $a/b=5$. The material is a commercially available dielectric denoted as K-250 from TCI Ceramics with $\epsilon_{r}=250$ and $\text{tg}\delta=0.005$. The lattice constant is 45 mm.}
\end{figure}

The high symmetry of DRs also suggest the possibilty of designing a 3D isotropic left-handed medium by combining them with a connected network of metallic wires \cite{Silveirinha-2005}. Actually, the field confinement concept underlying DR theory can be also applied to the design of a fully dielectric and isotropic left-handed metamaterial by substituting the metallic wires by dielectric rods (DRos). To keep the explanation as simple as possible we will stick to the parallel plate waveguide model of the cubic wire medium \cite[Sec. 2.2.2]{Marques}. In the frame of this model, the unit cell of such medium is equivalent to the circuit depicted in Fig. 5, where  $L/p = \mu _0$ is the per unit length inductance of ``vacuum'', $C/p  = \varepsilon$ is the per unit length capacitance of ``vacuum'', $Z_{\text{w}}$ is the impedance of a wire section of length $p$, and $p$ is the lattice constant. In metallic wire medium the wire exhibits an inductance that can be approximated as \cite{Silveirinha-2005}
\begin{equation}
  \label{eq9}
Z_{\text{w}}  = \frac{{j\omega \mu _0 p}}
{{2\pi }}\left[ {\ln \left( {\frac{p}
{{2\pi r_w }}} \right) + 0.5275} \right],
\end{equation}
where $r_{\text{w}}$ is the wire radius. However, as shown in the beginning of this paper, if the wires are substituted by  high permittivity DRos, they also exhibit an additional capacitance and resistance \reff{eq5} wich result in the impedance
\begin{equation}
  \label{eq10}
Z = \frac{p}
{{j\omega \varepsilon \pi r_{\text{w}}^2 }}
\end{equation}
which must be added to the impedance (\ref{eq9}). Now, the resulting effective permittivity is
\begin{equation}
  \label{eq11}
\varepsilon _{\text{rw}}  = 1 - \frac{1}
{{\frac{{\omega ^2 }}
{{\omega _{\text{p}}^2 }} - \frac{{\left( {p/r_{\text{w}} } \right)^2 }}
{{\varepsilon _{\text{r}} \pi }}}}, 
\end{equation}
where
\begin{equation}
  \label{eq12}
\omega _{\text{p}}^2  = \frac{1}
{{\frac{{\varepsilon _0 \mu _0 p^2 }}
{{2\pi }}\left[ {\ln \left( {\frac{p}
{{2\pi r_{\text{w}} }}} \right) + 0.5275} \right]}} 
\end{equation}
is the plasma frequency of the ordinary metallic wire medium. From \reff{eq11} it can be seen that the presence of the additional wire capacitance shifts the resonance of the metallic wire medium from zero frequency to the higher frequency $\omega_{\text{r}} = p w_{\text{p}} / \left( r_{\text{w}} \sqrt{\epsilon_r \pi} \right)$. The permittivity \reff{eq11} is plotted in Fig. 6 for some realistic structural parameters. 

\begin{figure}
\centering
\includegraphics[width=0.4\columnwidth]{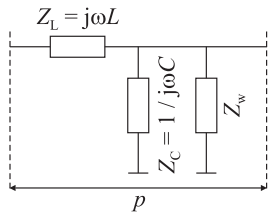}
\caption{\label{Fig5} Equivalent circuit model of the unit cell of the connected wire medium with cubic lattice of periodicity $p$.}
\end{figure}
\begin{figure}
\centering
\includegraphics[width=0.9\columnwidth]{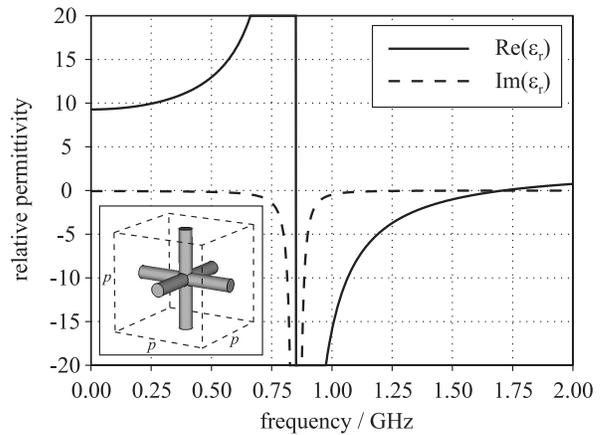}
\caption{\label{Fig6} Effective permittivity of a connected dielectric wire medium.  Geometrical parameters are $p = 45$ mm and $r_{\text{w}}= p / 15$. The material is a commercially available dielectric $\text{SrTiO}_3$ with $\epsilon_{r}=500$ and $\text{tg}\delta=0.01$. The unit cell is shown in the inset.}
\end{figure}

By combining the DR and DRo media of Figs. 4 and 6, the isotropic DR + DRo medium with the unit cell depicted in the inset of Fig. 7 is obtained. The band diagram along the $\Gamma-X$ direction of this medium calculated using CST Microwave Studio is shown in Fig. 7, along with the band diagram for the DR medium only. A backward-wave passband can be clearly observed for the DR + DRo medium. This passband appears quite approximately at the negative permeability frequency band of the DR medium shown in Fig. 4. The forward passband for the DR + DRo medium located at lower frequencies coincides almost exactly with a similar passband for the DRo medium alone, which corresponds to the region of positive dielectric permittivity below the resonance frequency $\omega_{\text{r}}$. The small numerical discrepancies in the location of the passbands and stopbands between Fig. 7, and Figs. 4 and 6 can be attributed to the effect of spatial dipersion, not consdered in our analytical model. It should be mentioned, however, that these effects could be included in the model using the general formalism developed in \cite{Baena-2008}.
\begin{figure}
\centering
\includegraphics[width=0.9\columnwidth]{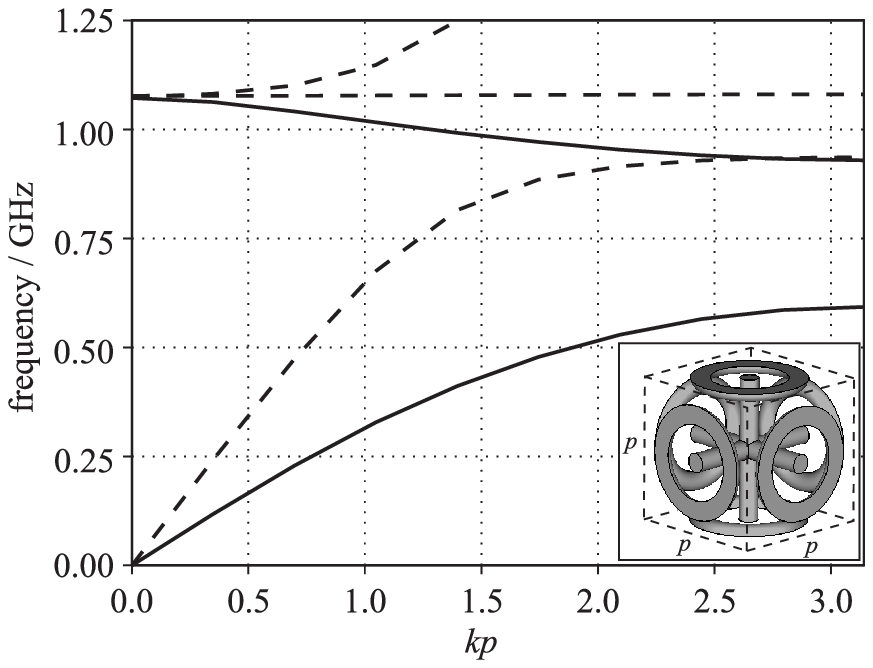}
\caption{\label{Fig7} Band diagram of the DR + DRo medium (solid lines) with unit cell shown in the inset and band diagram of the DR medium alone (dashed lines). Structural parameters are the same as in Fig. 4 and Fig. 6. Note that the flat longitudinal mode in DR band diagram will also appear in the combined structure, but has been omitted to keep the lucidity of the figure.}
\end{figure}

An interesting propery of DRs is compatibility with homogeneous negative permittivity media for left-handed media design. It is well known \cite{Pokrovsky-2002,Marques/Smith-2004,Marques} that mixing two composites of negative permittivity and permeability do not ensure a left-handed behavior. In particular, an array of SRRs immersed in a homogeneous negative permittivity medium do not provide a left-handed metamaterial, because the distributed capacitance along the SRR becomes negative (that is, in fact inductive) due to the negative permittivity of the host medium \cite{Marques}. However, the capacitance of the DR (\ref{eq5}) does not depend on the permittivity of the surrounding medium. Therefore, it can be guessed that the effective magnetic permeability of a DR arrangement will not change if the composite is immersed in a host medium of negative dielectric permittivity, thus resulting in a left-handed metamaterial. In order to show this property of DR composites, the transmittance through a five unit cell thick slab of the DR medium of Fig. 4, immersed in a plasma-like medium of permittivity $\varepsilon / \varepsilon_0 = 1-\omega_p^2/(\omega(\omega-\text{j} f_{\text{c}}))$, has been computed using CST Microwave Studio. The result is shown in Fig. 8. A clear passband in the frequency band of negative permeability of Fig. 4 can be observed, in agreement with the above hypothesis. Apart from this left-handed band, the structure also exhibits a right-handed passband at lower frequencies, which results from the plasmonic resonances supported by the ring for high values of the negative permittivity of the host medium. The ripples observed in both passbands are Fabri-Perrot resonances due to the finite width of the sample.

Apart from gaseous hot plasmas, low valued negative permittivity is present in metals and in semiconductors near its plasma frequency. Plasma frequency of metals is beyond the optical range, where achieving the large permittivities needed for the proposed DR designs seems almost impossible. However, some semiconductors such as InSb \cite{Gomez-Rivas-2005} and others, present plasma frequencies and relatively low losses in the terahertz range, where the high dielectric constants needed for DR media design can be easily achieved. Therefore, a semiconductor matrix with DR inclusions seems to be a promising structure for left-handed media design at THz frequencies.

\begin{figure}
\centering
\includegraphics[width=0.9\columnwidth]{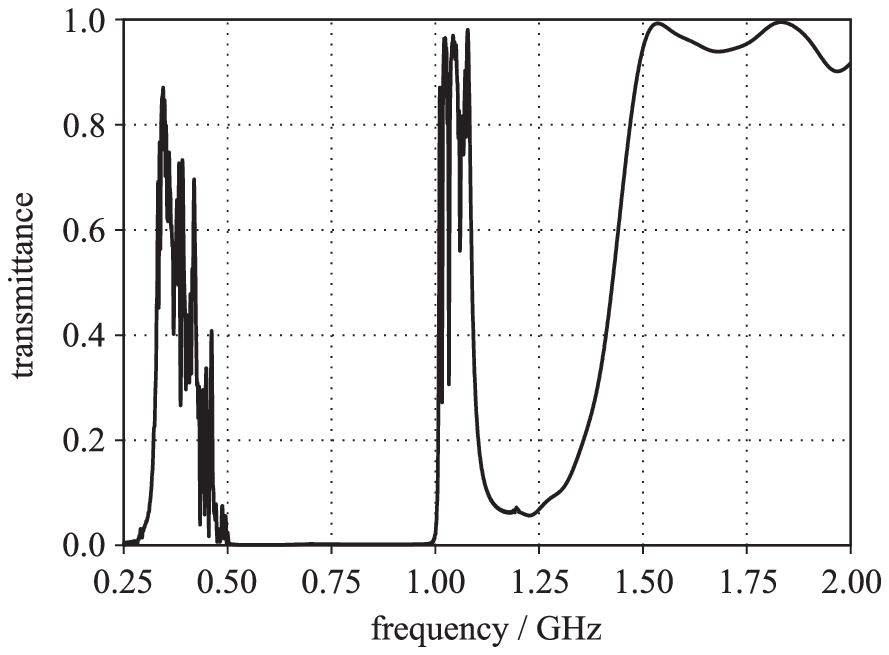}
\caption{\label{Fig8} Transmittance through five unit cells thick slab of DRs combined with homogeneous plasma. The plasma frequency has been set to $\omega_{\text{p}}=2 \pi 10^9 \sqrt{2}$  $s^{-1}$ and colision frequency to $f_{\text{c}} = \omega_{\text{p}} / 1000$.}
\end{figure}

In summary, it has been shown that a high permittivity DR can resonate at microwave and terahertz frequencies, providing strong artificial magnetism with a frequency band of effective negative permeability one order of magnitude larger than other previously proposed dielectric designs. By combining these DRs with DRos or with a plasma-like host medium, a left-handed behavior can be achieved. Combinations of DRs with DRos may be useful for the design of left-handed metamaterials when interactions with external low frequency magnetic fields should be minimized. A composite made of DR inclusions in a host semiconducting media near its plasma frequency appears as a promising alternative for the design of left-handed media at terahertz frequencies.

\begin{acknowledgments}
This work has been supported by the Spanish Ministerio de Educaci\'on y Ciencia and
European Union FEDER funds (projects TEC2007-65376, TEC2007-68013-C02-01, and
CSD2008-00066), and by Junta de Andaluc\'\i a (project TIC-253). L. Jelinek also thanks for 
the support of the Czech Grant Agency (project no. 102/08/0314).
\end{acknowledgments}

\end{document}